%% file: iclr2021_conference.tex
\documentclass{article} 
\usepackage{iclr2021_conference,times}

\input{math_commands.tex}

\usepackage{hyperref}
\usepackage{url}
\usepackage[utf8]{inputenc} 
\usepackage[T1]{fontenc}    
\usepackage{booktabs}       
\usepackage{amsfonts}       
\usepackage{subfig}
\usepackage{nicefrac}       
\usepackage{microtype}      
\usepackage{graphicx}       
\usepackage[version=4]{mhchem}         
\usepackage{booktabs}
\usepackage{multirow}
\usepackage{makecell}
\usepackage{threeparttable}
\usepackage{natbib}

\title{Investigating Ground-level Ozone Formation: A Case Study in Taiwan}

\author{
  Yu-Wen Chen$^{1}$, Sourav Medya$^2$, Yi-Chun Chen$^1$ \\
  $^1$Acdemia Sinica, Taiwan \hspace{0.5cm} $^2$Northwestern University, USA \\
 $^1$\texttt{\{yuwenchen,yichunchen\}@gate.sinica.edu.tw} \\
 $^2$\texttt{sourav.medya@kellogg.northwestern.edu } 
}

\iclrfinalcopy 
\begin{document}

\maketitle
\lhead{AI: Modeling Oceans and Climate Change Workshop at ICLR 2021}

\input{0_abstract}

\input{1_intro}
\input{2_method}

\input{3_results}

\input{4_conclusion}

\bibliography{iclr2021}
\bibliographystyle{iclr2019_conference}



\end{document}

%% file: math_commands.tex
\usepackage{amsmath,amsfonts,bm}

\def\eqref#1{equation~\ref{#1}}

\def\1{\bm{1}}

\DeclareMathAlphabet{\mathsfit}{\encodingdefault}{\sfdefault}{m}{sl}
\SetMathAlphabet{\mathsfit}{bold}{\encodingdefault}{\sfdefault}{bx}{n}



%% file: 0_abstract.tex
\begin{abstract}
  Tropospheric ozone (\ce{O3}) is a greenhouse gas which can absorb heat and make the weather even hotter during extreme heatwaves. Besides, it is an influential ground-level air pollutant which can severely damage the environment. Thus evaluating the importance of various factors related to the \ce{O3} formation process is essential. However, \ce{O3} simulated by the available climate models exhibits large variance in different places, indicating the insufficiency of models in explaining the \ce{O3} formation process correctly.
  In this paper, we aim to identify and understand the impact of various factors on \ce{O3} formation and predict the \ce{O3} concentrations under different pollution-reduced and climate change scenarios. We employ six supervised methods to estimate the observed \ce{O3} using fourteen meteorological and chemical variables. We find that the deep neural network (DNN) and long short-term memory (LSTM) based models can predict \ce{O3} concentrations accurately. We also demonstrate the importance of several variables in this prediction task. The results suggest that while Nitrogen Oxides negatively contributes to predicting \ce{O3}, solar radiation makes a significantly positive contribution. Furthermore, we apply our two best models on \ce{O3} prediction under different global warming and pollution reduction scenarios to improve the policy-making decisions in the \ce{O3} reduction.
  
\end{abstract}

%% file: 1_intro.tex
\section{Introduction}
Ozone (\ce{O3}) plays an essential role in the stratosphere to prevent organisms in the biosphere from exposing to excessive ultraviolet (UV) rays \citep{seinfeld2016}. However, it is also a greenhouse gas and a severe air pollutant at the ground-level. Ground-level \ce{O3} can absorb longwave radiation from the earth, further shifting the radiation balance and even heating the surrounding atmosphere \citep{Stevenson2013}. High concentrations of the ground-level \ce{O3} can also severely damage the ecological community. For instance, $4-15\%$ of global wheat yields are lost because of \ce{O3} pollution \citep{ain2017}. Therefore, the Environmental Protection Agency (EPA) of the United States set a National Ambient Air Quality Standards (NAAQS) for six principal pollutants including ground-level \ce{O3}. According to the latest 2015 NAAQS, the standard for ground-level \ce{O3} is 0.070 ppm for an eight-hour average. Considering that tropospheric \ce{O3} is produced through complicated reactions, understanding the importance of different variables and their interactions that produce \ce{O3} is necessary. However, as an obvious model-observation disparity of tropospheric \ce{O3} still exists in current global-scale chemical models developed based on theoretical studies \citep{young2018}, analyzing its formation with new data-driven methods becomes essential.

Several popular machine learning algorithms have been applied in the real-time prediction and down-scaling of \ce{O3} concentration ~\citep{esl2019,gre2019}. These methods are also used to simplify the \ce{O3} prediction process in climate models and reduce the computational expense of fully interactive atmospheric chemistry schemes \citep{nowack2018}. However, most of them focus only on the prediction task and ignore the comparison of the importance of the features with earlier theoretical studies. It is essential to understand the importance of the factors in the complex
\ce{O3} formation process to help improve policy-making and progress towards a healthier environment. In this study, we predict ground-level \ce{O3} with different machine learning models and measure the importance of several factors involved in \ce{O3} formation.

The availability of enormous data such as satellite observation images and uninterrupted surface measurements helps in understanding tropospheric \ce{O3} formation. The present modern techniques (e.g., deep learning methods) are also compatible with large-scale data and highly effective in making predictions. In this paper, we utilize large scale data and modern deep learning techniques to understand \ce{O3} formation. The meteorological parameters and the concentration of pollutants are adequate for ground-level \ce{O3} prediction. We use fourteen such variables in our analysis. The observed \ce{O3} is regarded as true values for the prediction. We aim to learn a prediction function $f$ that takes these fourteen variables as input features ($\mathbb{X}$) and predict the value ($y$) of the observed \ce{O3}. Our main contributions are as follows:
\begin{itemize}
    \item We collect a large dataset on observed \ce{O3} and corresponding important weather factors. We build several supervised learning methods to accurately predict \ce{O3} concentrations.
    \item We demonstrate the importance of several factors (variables) in this prediction task by applying two well-known frameworks for identifying feature importance.
    \item We apply our two best models under different global warming and pollution reduction scenarios to improve the policy-making decisions in the \ce{O3} reduction.
\end{itemize}

%% file: 2_method.tex
\section{Data and Methods}
\label{sec:method}

\begin{figure*}[ht]
    \centering
    \vspace{-3mm}
    \subfloat[EPA stations distribution]{\includegraphics[width=0.31\textwidth]{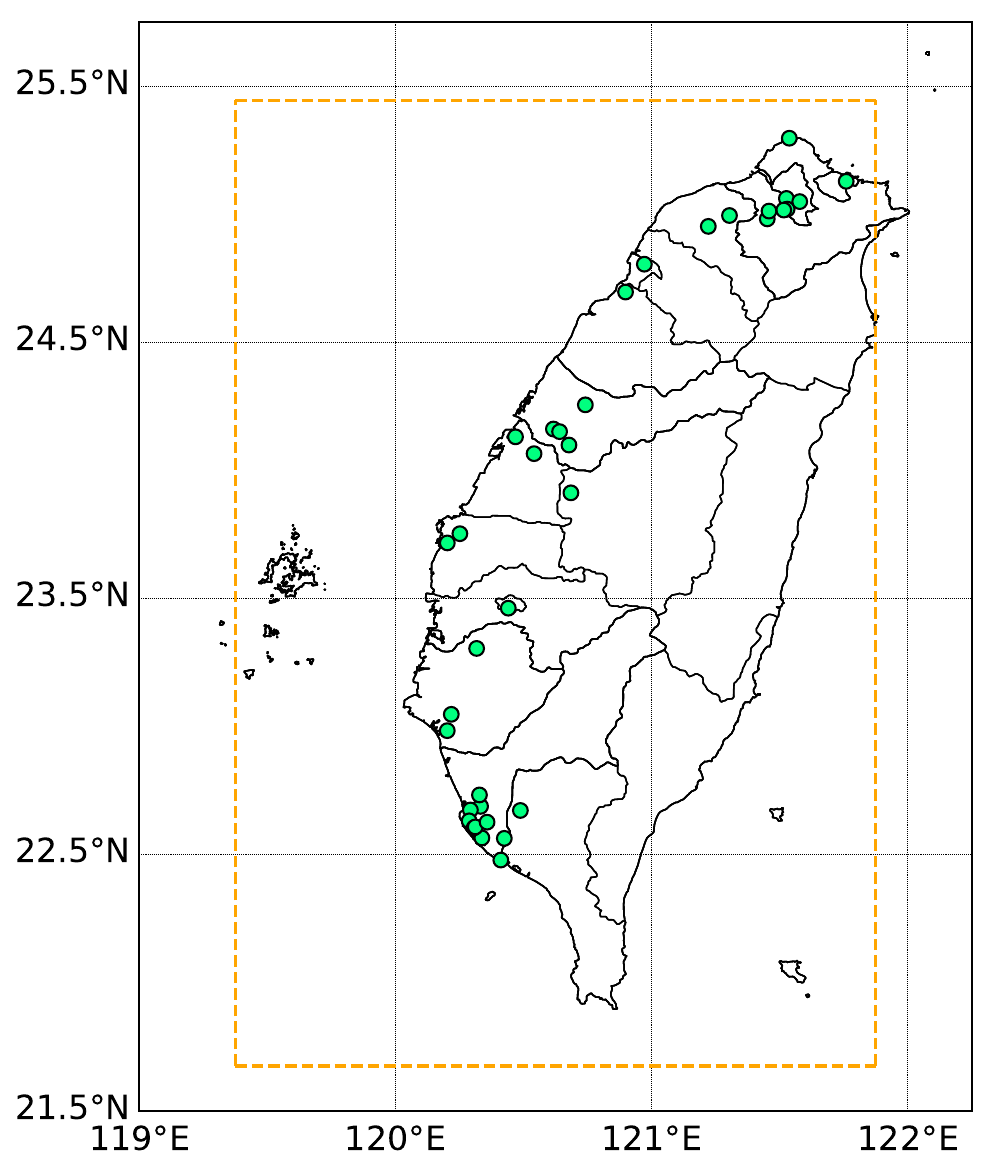} \label{fig:epa_distribution}}
    \subfloat[Experiment procedure]{\includegraphics[width=0.54\textwidth]{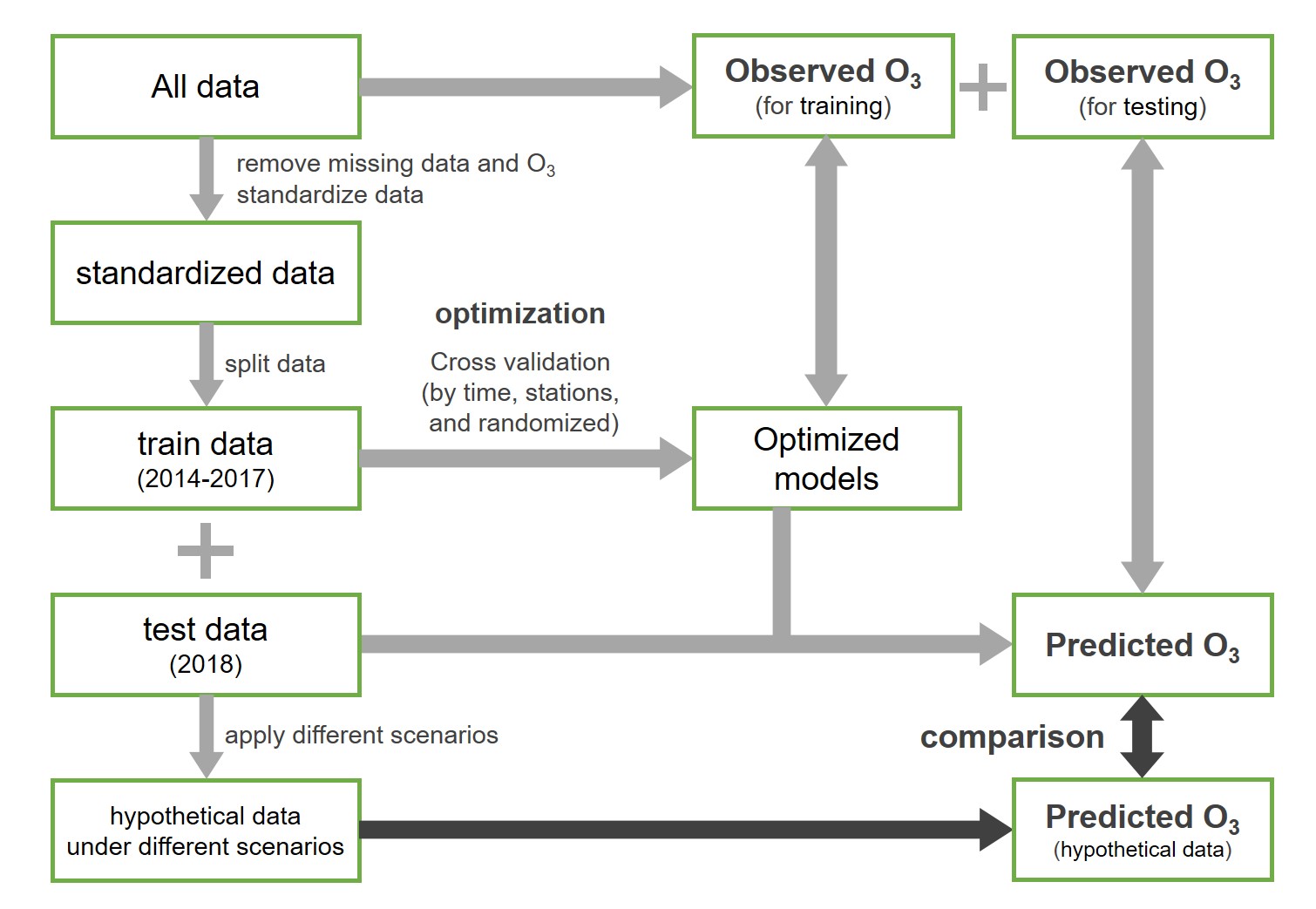} \label{fig:procedures}}
    
    \caption{(a) The distribution of the EPA stations in Taiwan. The orange frame is the area that we analyze from the CESM2 model, which covers most Taiwan island and outlying islands. (b) The scheme of the experiment procedures.}
   \vspace{-2mm} 
\end{figure*}

\paragraph{Dataset}
The dataset contains $14$ variables of three different types and a total of $3,204,710$ hourly data points observed during the span of $2014-2018$. We combine consecutive eight hourly data points which does not include any missing value to generate a new 2-dimension dataset (eight hours of 14 variables). We use data observed in $2014-2017$ ($655,850$ points) and in $2018$ ($225,478$ points) for training and testing respectively. The three different types of variables are as follows. \textbf{(1) The in-site measurements}: These include $12$ variables measured every hour at $36$ surface stations arranged by the Environmental Protection Agency (EPA) of Taiwan, as shown in Figure \ref{fig:epa_distribution}. \textbf{(2) The derived variable}: These contains water vapor mixing ratio converted from the previous EPA information. \textbf{(3) The observations from remote sensor:} These are the surface downward solar radiation (rsds) data inverted from the absorption data of Himawari 8, a Japan’s geostationary meteorological satellite, by the Central Weather Bureau and National Science and Technology Center for Disaster Reduction of Taiwan \citep{Himawari2016}. Table \ref{tab:training-testing-dataset-info} presents the details of each variable. The values of these inputs (independent) variables are observed hourly.

\begin{table}[hb]

\centering
\small
\begin{tabular}{@{}lll@{}}
\toprule
\textbf{Variable}                            & \textbf{Unit}    & \textbf{Data source}                          \\ \midrule
Air temperature (T)                          & $^{\circ}$C      & Taiwan EPA                                    \\
Wind speed (WS)                             & m/s              & Taiwan EPA                                    \\
Wind direction (WDIR)                       & degree           & Taiwan EPA                                    \\
Relative humidity (RH)                       & \%               & Taiwan EPA                                    \\
Water vapor mixing ratio ($\textrm e_w$)         & g/kg             & Convert from RH and T                         \\
Surface downward solar radiation (rsds) & $\textrm W/m^2$ & Calculate from Himawari 8 observation \\
Nitric oxide (NO)                            & ppb              & Taiwan EPA                                    \\
Nitrogen dioxide (\ce{NO2}) & ppb              & Taiwan EPA                                    \\
Carbon monoxide (CO)                         & ppm              & Taiwan EPA                                    \\
Methane (\ce{CH4})                         & ppm              & Taiwan EPA                                    \\
Non-methane Hydrocarbon (NMHC)           & ppm              & Taiwan EPA                                    \\
Sulfur dioxide (\ce{SO2})   & ppb              & Taiwan EPA                                    \\
$\rm PM_{2.5}$                               & $\rm \mu g/ m^3$ & Taiwan EPA                                    \\
$\rm PM_{10}$                                & $\rm \mu g/ m^3$ & Taiwan EPA                                    \\
Ozone (\ce{O3})             & ppb              & Taiwan EPA                                    \\ \bottomrule

\end{tabular}
\caption{The variables in the dataset: Observed \ce{O3} is regarded as true values for predictions. $WS$ is measured in meter per second (m/s). $WDIR$ is recorded in degrees from 0 to 360 with 0 as north. $T$ is measured in Celsius ($^{\circ}$C). The $e_w$ is converted to gram per kilogram air (g/kg). Trace gases are measured in either parts-per million (ppm) or parts-per billion (ppb). Particulate matters are recorded in microgram per cubic meter air ($\rm \mu g/ m^3$).}
\label{tab:training-testing-dataset-info}
\end{table}

In addition to the observed data, monthly historical simulation (2000-2014) \citep{CESM2_historical_2019} and future projection (2015-2100) \citep{CESM2_ssp126_2019, CESM2_ssp245_2019, CESM2_ssp370_2019, CESM2_ssp585_2019} from CESM2 are used to evaluate the future trend of \ce{O3}. CESM2 is a global climate model developed by the US National Center for Atmospheric Research \citep{CESM2_2020}. The variables including temperature, relative humidity and water vapor content are analyzed in the form of an area average (longitude from 119.375 $^{\circ}$E to 121.875 $^{\circ}$E and latitude from 21.675 $^{\circ}$N to 25.445 $^{\circ}$N, as displayed in Fig. \ref{fig:epa_distribution}).

\paragraph{Methods}
 We formulate our problem as a regression problem and use six different algorithms: linear regression (LR), random forest (RF), optimized distributed gradient boosting model (XGBoost), convolution neural network (CNN), deep neural network (DNN), and long short-term memory model (LSTM). We aim to predict the eight-hour average observed ground-level ozone. The DNN model consists of five hidden layers with 16 nodes each. The CNN model is made up of 2 convolution layers of 32 nodes with a 3x3 window, a max pooling, a flattening, and a fully-connected layer. 20\% data are dropped out after the first convolution layer and the max pooling layer individually. The LSTM model consists of two LSTM layers with 25 nodes each and a fully-connected layer. The previously described consecutive 8-hour 14 variables are reshaped to the 1-dimensional input data for models including LR, RF, XGBoost, and DNN. 
The consecutive 8-hour 14 variables is prepared as the 2-dimension input data for the CNN and LSTM models. We describe the entire experimental setup in Figure \ref{fig:procedures}.

%% file: 3_results.tex
\section{Experimental Results}
\label{sec:results}

We demonstrate three types of results. First, we describe the performance of all the proposed models via out of sample tests. Second, we present the importance of the input variables in predicting \ce{O3}. Third, we evaluate the impact of climate change and pollution on the ground level \ce{O3} and explain how these results would help in better policy making.  

\subsection{Model performance comparison}
\begin{figure*}[t]
\vspace{-4mm}
    \centering
    \subfloat[Performance Results]{\includegraphics[width=0.32\textwidth]{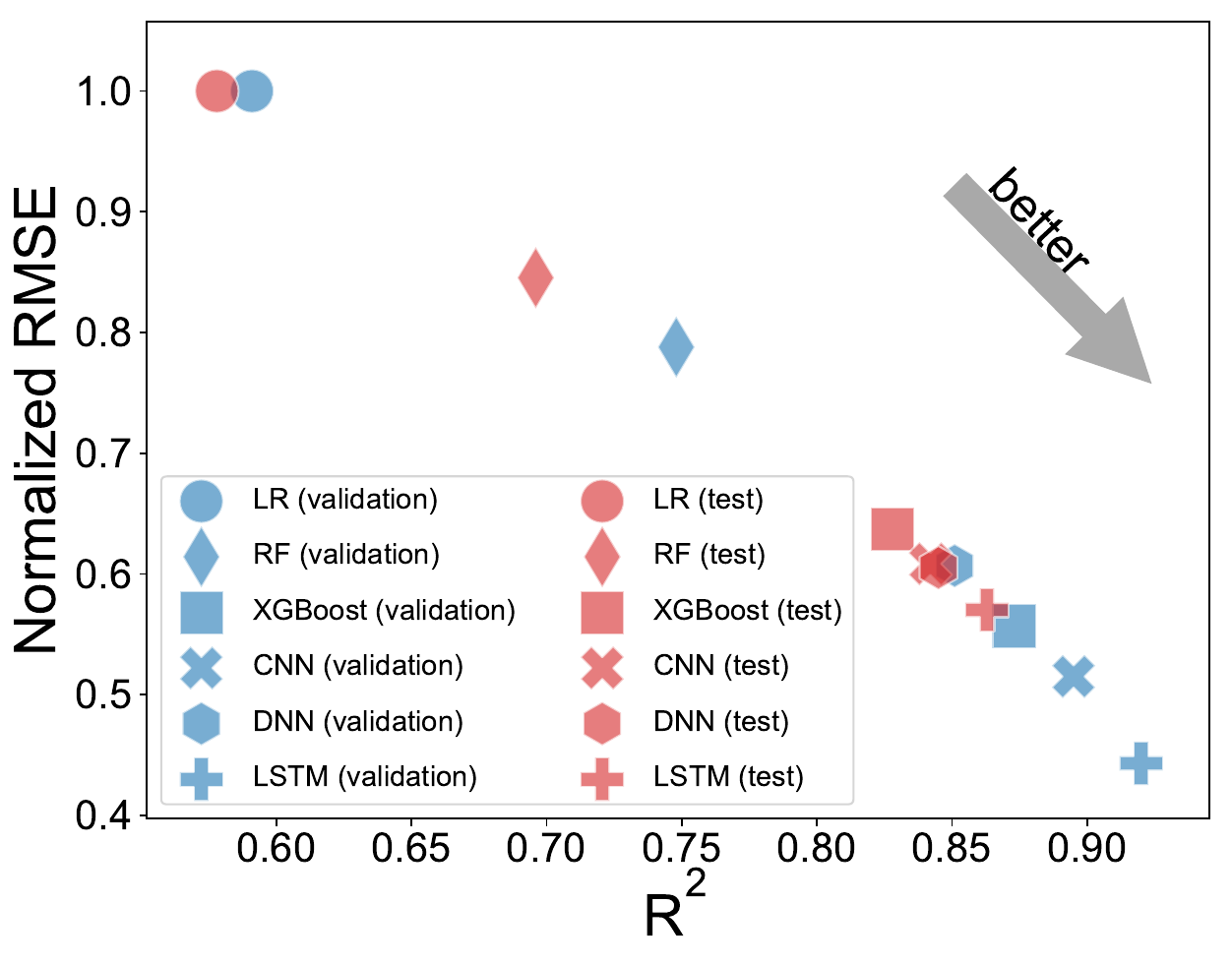} \label{fig:NRMSE-to-R2}}
    \subfloat[Predicted vs Observed]{\includegraphics[width=0.6\textwidth]{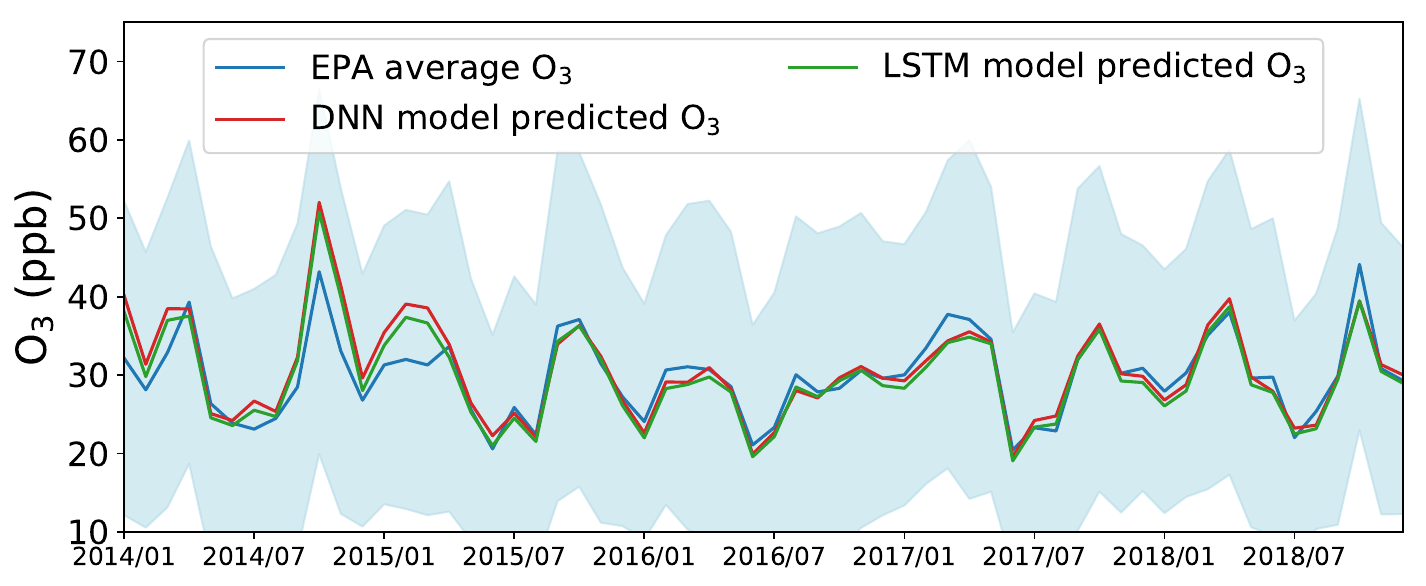}\label{fig:O3_epa_pred}}
    
    \caption{(a) The performance of six models on the validation and test data. (b) The EPA monthly average \ce{O3}, DNN model predicted \ce{O3} and LSTM model predicted \ce{O3} in 2014-2018. }
  \vspace{-4mm}  
\end{figure*}

We compare the performance of all six models in this experiment. The training and testing data are from the span of $2014-2017$ and $2018$ respectively. For validation, 10\% of the data in $2014-2017$ is chosen in three ways: i) \textbf{sample}: randomly 10\% selection from the entire data; ii) \textbf{station}: randomly selecting data from 10\% stations, iii) \textbf{date}: randomly selecting data from 10\% dates in each month. Note that the test data is always fixed and is from the year $2018$. The $\textrm R^2$ and root mean square error (RMSE) (\cite{gre2019}) between the model-predicted eight-hour average \ce{O3} and EPA measured eight-hour average \ce{O3} are used as the performance measures.
The results for all three types of validation methods and their corresponding test results are presented in Table \ref{tab:model_performance}. The LSTM and DNN are the best performing models with high $\textrm R^2$ ($>.84$) and low RMSE ($<6.65$) in predicting the observed \ce{O3}. Furthermore, Fig. \ref{fig:NRMSE-to-R2} presents the $\textrm R^2$ and normalized RMSE produced by all the models when the validation set is randomly 10\% selected from the entire data.
To further visualize the actual predictions of the DNN and LSTM models, we compare the values of predicted and observed \ce{O3} in Fig. \ref{fig:O3_epa_pred}. Note that the predicted \ce{O3} by the DNN and LSTM model is correlated with the observed ones and this justifies the good performance of the model. 

\begin{table}[t]
\centering
\small
\begin{tabular}{@{}cccrrrrrr@{}}
\toprule
\textbf{Data} &
\textbf{Division rule} &
   &
  \multicolumn{1}{c}{\textbf{LR}} &
  \multicolumn{1}{c}{\textbf{RF}} &
  \multicolumn{1}{c}{\textbf{XGBoost}} &
  \multicolumn{1}{c}{\textbf{CNN}} &
  \multicolumn{1}{c}{\textbf{DNN}} &
  \multicolumn{1}{c}{\textbf{LSTM}} \\ \midrule
\multirow{9}{*}{\makecell{Validation \\ data}} & \multirow{3}{*}{sample}   & slope & 0.592  & 0.710  & 0.872 & 0.856 & 0.872 & 0.923 \\
                                 &                          & $\textrm R^2$ & 0.591  & 0.748  & 0.873 & 0.895 & 0.872 & 0.920 \\
                                 &                          & RMSE  & 11.083 & 8.732 & 6.170 & 5.707 & 6.724 & 4.909 \\ \cmidrule{2-9}%
                                 & \multirow{3}{*}{station} & slope & 0.597  & 0.702  & 0.859 & 0.812 & 0.439 & 0.896 \\
                                 &                          & $\textrm R^2$ & 0.590  & 0.733  & 0.848 & 0.853 & 0.994 & 0.866 \\
                                 &                          & RMSE  & 11.196 & 8.942 & 6.828 & 6.763 & 4.148 & 6.424 \\ \cmidrule{2-9}%
                                 & \multirow{3}{*}{date}    & slope & 0.923  & 0.698  & 1.185 & 0.709 & 0.894 & 0.411 \\
                                 &                          & $\textrm R^2$ & 0.899  & 0.984  & 0.895 & 0.975 & 0.890 & 0.964 \\
                                 &                          & RMSE  & 4.708 & 4.646 & 4.574 & 4.130 & 5.740 & 5.055 \\ \specialrule{0.05pt}{1pt}{1pt} 
\multirow{9}{*}{\makecell{Test \\ data}}       & \multirow{3}{*}{sample}   & slope & 0.572  & 0.678  & 0.860 & 0.829 & \underline{0.883} & \textbf{0.903} \\
                                 &                          & $\textrm R^2$ & 0.578  & 0.696  & 0.828 & 0.842 & \underline{0.845} & \textbf{0.863} \\
                                 &                          & RMSE  & 10.855 & 9.176 & 6.916 & 6.603 & \underline{6.568} & \textbf{6.188} \\ \cmidrule{2-9}%
                                 & \multirow{3}{*}{station} & slope & 0.576  & 0.686  & 0.858 & 0.826 & \underline{0.875} & \textbf{0.896} \\
                                 &                          & $\textrm R^2$ & 0.577  & 0.696  & 0.829 & 0.840 & \underline{0.841} & \textbf{0.857} \\
                                 &                          & RMSE  & 10.861 & 9.198 & 6.906 & 6.641 & \underline{6.647} & \textbf{6.305} \\ \cmidrule{2-9}%
                                 & \multirow{3}{*}{date}    & slope & 0.571  & 0.678  & 0.862 & 0.820 & \underline{0.882} & \textbf{0.896} \\
                                 &                          & $\textrm R^2$ & 0.578  & 0.694  & 0.829 & 0.842 & \underline{0.845} & \textbf{0.860} \\
                                 &                          & RMSE  & 10.856 & 9.213 & 6.903 & 6.606 & \underline{6.573} & \textbf{6.257} \\ \bottomrule
\end{tabular}
\caption{The performance of each model on validation and test data. Three division methods to split data for training and validating are used: (1) randomly choosing 10\% samples for validation (2) randomly selecting data from 10\% stations (3) randomly selecting data from 10\% dates in each month. The slope, $\textrm R^2$, and RMSE for validation data are the average of results from 10-fold CV. Models which have the lowest RSME in 10-fold CV are utilized for test data evaluation. In the test data, the best two performances are in bold and underlined respectively.\label{tab:model_performance}}
\end{table}

\subsection{Importance of different factors in predicting \ce{O3}}

\begin{figure*}[t]
    \centering
    \vspace{-3mm}
    \subfloat[Permutation importance]{\includegraphics[width=0.242\textwidth]{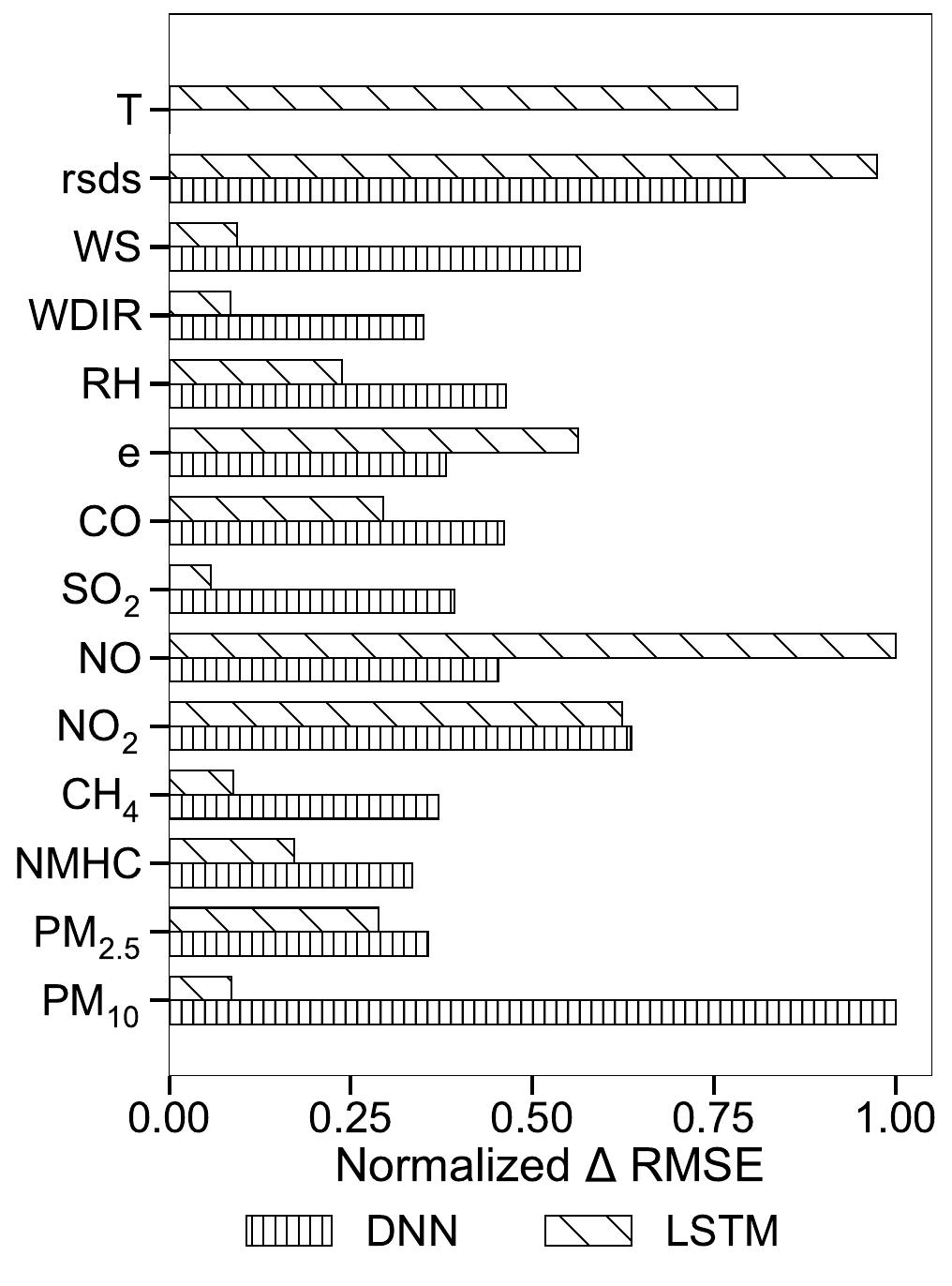} \label{fig:permutation_importance}}
    \subfloat[SHAP value of Variables]{\includegraphics[width=0.243\textwidth]{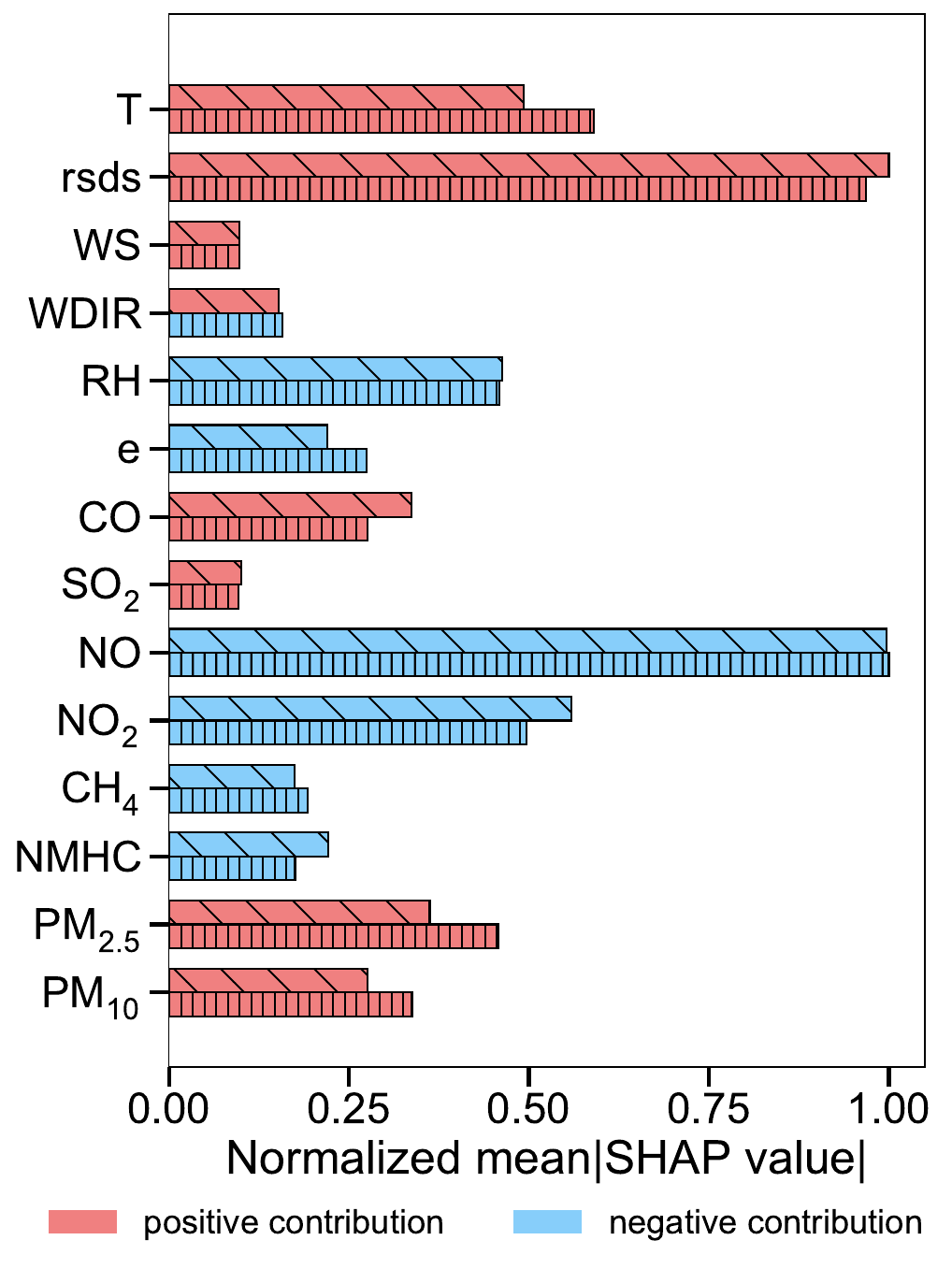} \label{fig:shap_importance}}
    \subfloat[Predicted vs Observed]{\includegraphics[width=0.4\textwidth]{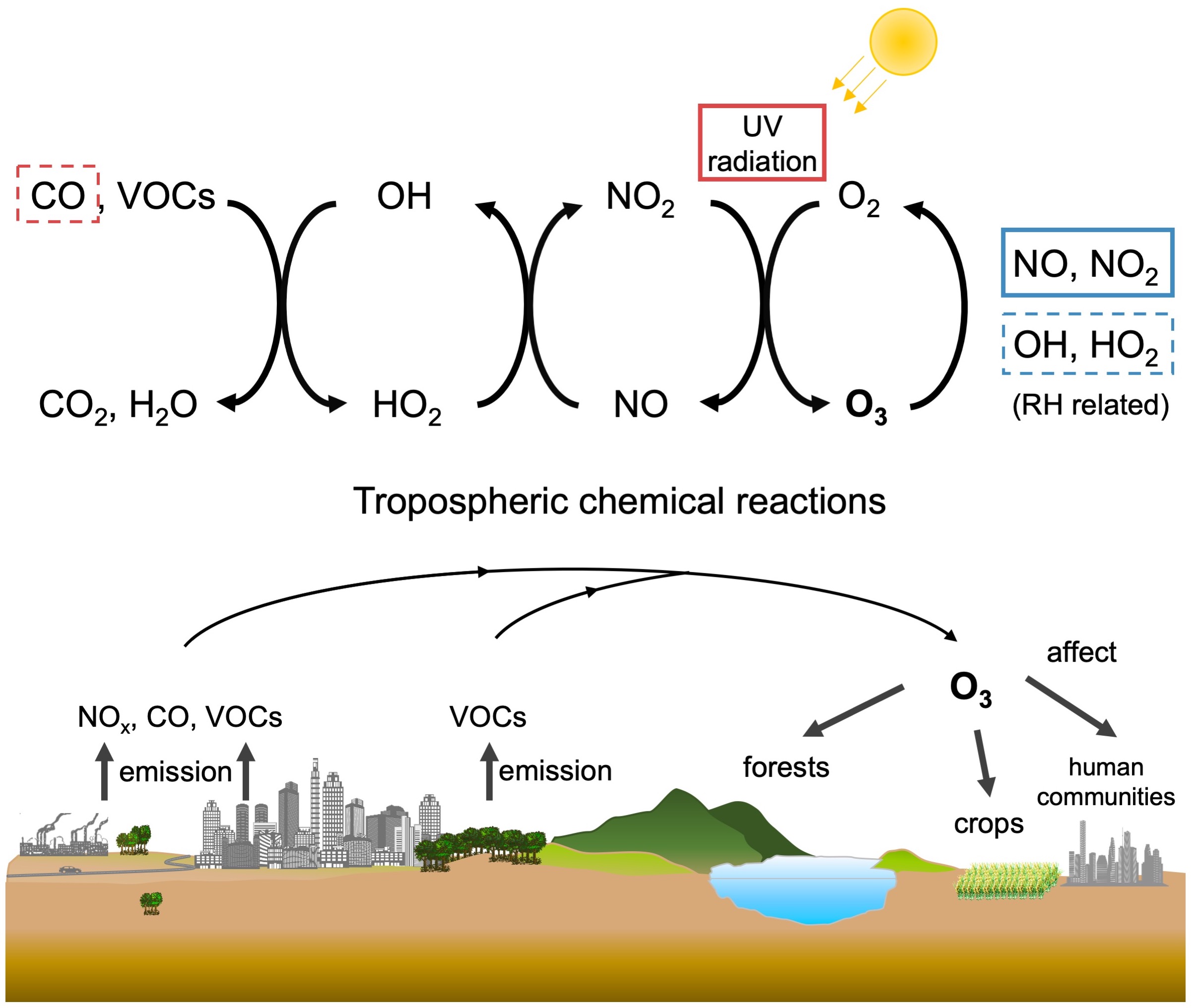}\label{fig:rxn_scheme}}

    \caption{Importance of variables given by the DNN and LSTM models via two techniques: (a) permutation importance \citep{alt2010} and (b) SHAP \citep{SHAP2017}. The SHAP values are the average values of all input hour. (c) Scheme of the simplified tropospheric \ce{O3} formation and decomposition: Anthropogenic trace gases (including \ce{NO}, \ce{NO2} CO, and VOCs) and natural trace gases (some VOCs, such as isoprene and terpenes) can affect the formation and decomposition of \ce{O3} directly or indirectly. CO and VOCs can react with OH radical, a significant atmospheric oxidant, to generate \ce{HO2} which will further convert NO to \ce{NO2}. Afterwards, \ce{NO2} can react with oxygen (\ce{O2}) to form \ce{O3} in the presence of UV radiation. The produced \ce{O3} might get decomposed due to the reaction with \ce{NOx} and \ce{HOx}. The generated \ce{O3} can severely impact the environment. The most important variables related to predict \ce{O3} are solar radiation and \ce{NOx} family, which contribute positively and negatively to the predicted \ce{O3}. The minor essential factors are humidity and CO, which contribute negatively and positively to predict \ce{O3}. }
   \vspace{-2mm} 
\end{figure*}

The tropospheric \ce{O3} formation process is complex and is influenced by many variables. One of our main goals is to understand the influence of individual variables on \ce{O3} prediction. Thus, we perform a permutation importance \citep{alt2010} study with the two best performing models DNN and LSTM. The idea is to make one feature randomly unavailable and then compute the drop in model's performance. Note that this method is also model agnostic. We measure the increase in RMSE ($\rm \Delta $ RMSE) as the drop in model's performance. The results shown in Fig. \ref{fig:permutation_importance} suggest that solar radiation is the most significant variable among uncontrollable variables identified by both models. On the other hand, the DNN model emphasizes the significance of $\rm PM_{10}$ while the LSTM model presents the importance of nitrogen monoxide (\ce{NO}) among variables that are related to human activities and controllable. These results also show that \ce{NO2} is another major important anthropogenic variable in \ce{O3} prediction. The similarity of two analyses signifies the influence of solar radiation and \ce{NO2} in the prediction of \ce{O3}. Other components having high importance according to the permutation analyses include relative humidity (RH), carbon monoxide (CO), and air temperature. Note that \ce{NO}, \ce{NO2}, CO and $\rm PM_{10}$ can be reduced or controlled with better policies such as reduction in fuel combustion and increasing usage of electric vehicles.

We further recognize the nature (positive or negative) of the contribution of each variables in predicting \ce{O3} via a recent technique. We employ SHapley Additive exPlanations (SHAP) analysis \citep{SHAP2017} on DNN and LSTM models. The idea is to use a game theoretic approach called Shapley Values to compute contribution of each individual variable. As displayed in Fig. \ref{fig:shap_importance}, the results indicate that while NO and \ce{NO2} have strong negative impact, the radiation has extremely positive influence on model-predicted \ce{O3}.

These results shed some light on the important mechanisms related to \ce{O3} formation, as further illustrated in Figure \ref{fig:rxn_scheme}. An interesting observation is that the negative contribution of \ce{NO} and \ce{NO2} might imply that ground-level \ce{O3} in Taiwan is mainly constrained by the "\ce{NOx}-saturated" condition \citep{sil1999}. In other words, reducing the emission of the volatile organic compounds (VOCs), such as benzene, formaldehyde, methanol, and isoprene, might be more efficient than \ce{NOx} for curtailing the surface \ce{O3} concentration in Taiwan. However, the slightly negative contribution of \ce{CH4} and non-methane hydrocarbon indicate that reducing these VOCs might not effectively reduce ground-level \ce{O3} neither. As a result, the contradiction to the traditional theory could provide a hint for further interesting research directions towards the unrevealed mechanisms of \ce{O3} formation. Besides, the high significance of $\rm PM_{10}$ in the permutation analysis of DNN model could not only present the high correlation between $\rm PM_{10}$ and \ce{O3} but also be a notion to further explore the possible mechanism for \ce{O3} formation on the surface of $\rm PM_{10}$.

\subsection{Evaluation of the impact of climate change and pollution reduction on ground-level \ce{O3} concentration}

Accurate \ce{O3} projection is an important task to help in improving environment related policies that include pollutant emission reduction and damage mitigation. In particular, monitoring and evaluating the impact of \ce{O3} on agricultural crops are important since agricultural production losses might cause food crisis and even famine around the world. Here we aim to perform the \ce{O3} prediction for different scenarios by applying our proposed DNN and LSTM models. The DNN and LSTM models are able to predict the monthly average data \ce{O3} concentration quite accurately (Fig. \ref{fig:O3_epa_pred}). Thus, we apply them for analyzing the impact of the pollution reduction and climate change on predicted \ce{O3}. 

\paragraph{Climate change scenarios}
''Shared Socioeconomic Pathways'' project four different climate change scenarios that are referred as ssp126, ssp245, ssp370, and ssp585 \citep{Neill-2016}. The ssp126 scenario presumes people to "take the green road" that the world shifts gradually toward a more sustainable path. The ssp245 scenario is a "middle of the road" that the world nearly follows their historical patterns. The ssp370 scenario assumes a “regional rivalry” that weak action is taken on mitigating climate and reducing air pollutant emissions. The ssp585 suspects that the world chooses to accelerate their growth in economic output and energy use. The model simulations based on these four scenarios show that surface temperature over Taiwan is expected to raise 1.0, 1.6, 2.5, and 3.7 $^{\circ}$C in by the end of 2100 (2091-2100) compared to 2014-2018 (the period this study focuses). In addition, the water vapor content is supposed to increase 5\%, 11\%, 17\% and 24\% in different scenarios. The relative humidity has less change compared to near-surface temperature and water vapor content, as presented in Table \ref{tab:cc_variables_change}. We apply our DNN and LSTM models on all of these four scenarios.


\begin{table}[]
\centering
\begin{tabular}{ccccc}
\hline

 & $\Delta$T ($^{\circ}$C) & $\Delta$e (\%) & $\Delta$RH (\%) \\ \hline
ssp126 & 1.0 & 5 & -0.2 \\
ssp245 & 1.6 & 11 & 0.4 \\
ssp370 & 2.5 & 17 & -0.3 \\
ssp585 & 3.7 & 24 & -0.8 \\ \hline
\end{tabular}
\caption{The change of three major factors (variables) in CESM2 \citep{CESM2_2020} under ssp126, ssp245, ssp370 and ssp585 scenarios by the end of 21 century (2091-2100) compared to the study period (2014-2018) over Taiwan. $\Delta$T is the average temperature change in degrees Celsius. $\Delta$e and $\Delta$RH are the water vapor change and relative humidity change in percentage, respectively.}
\label{tab:cc_variables_change}
\end{table}

\begin{figure*}[ht]
    \centering
    \vspace{-3mm}
    \subfloat[Climate change scenarios]{\includegraphics[width=0.445\textwidth]{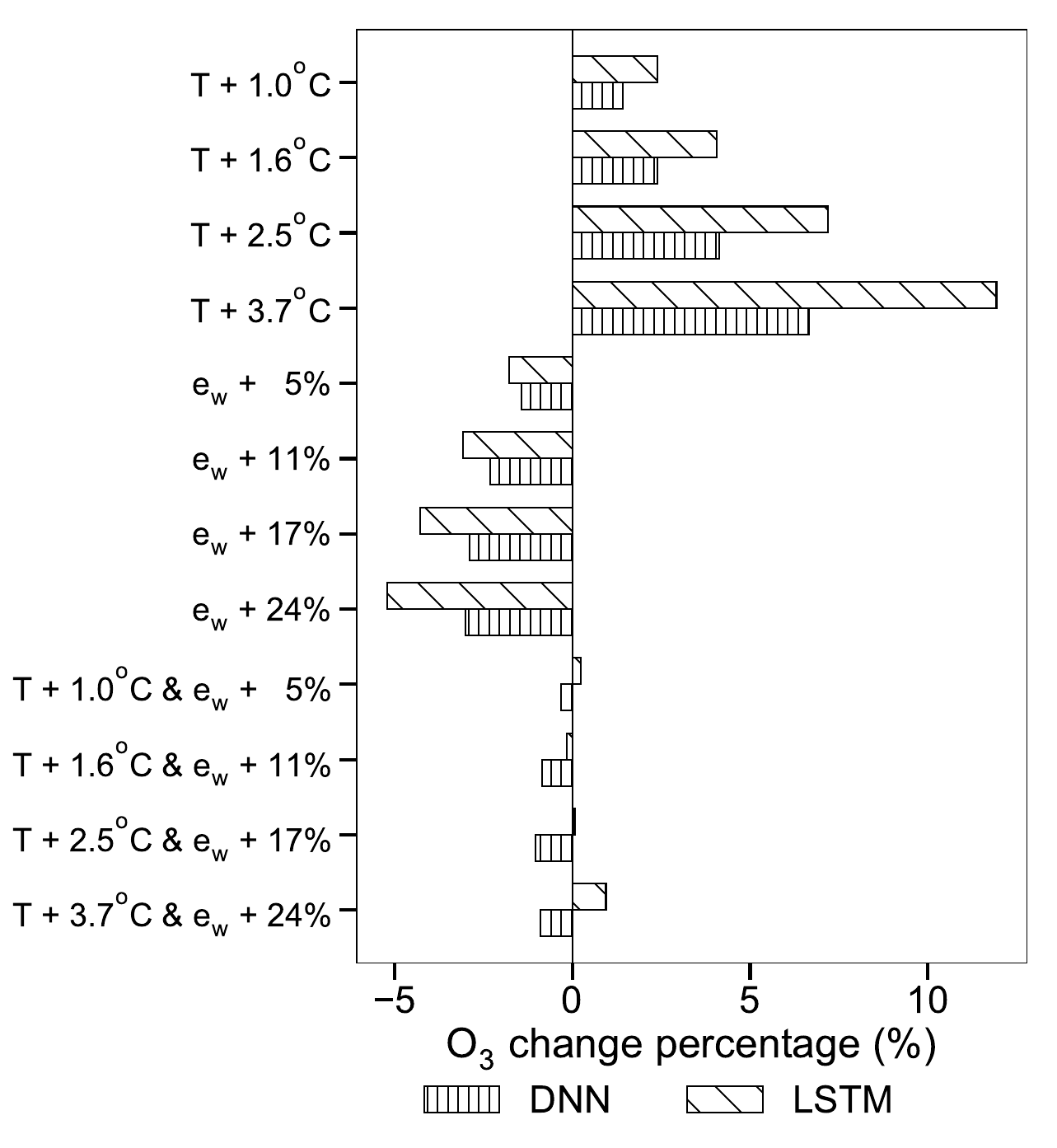} \label{fig:cc_simulation}}
    \subfloat[Pollution controlling scenarios]{\includegraphics[width=0.405\textwidth]{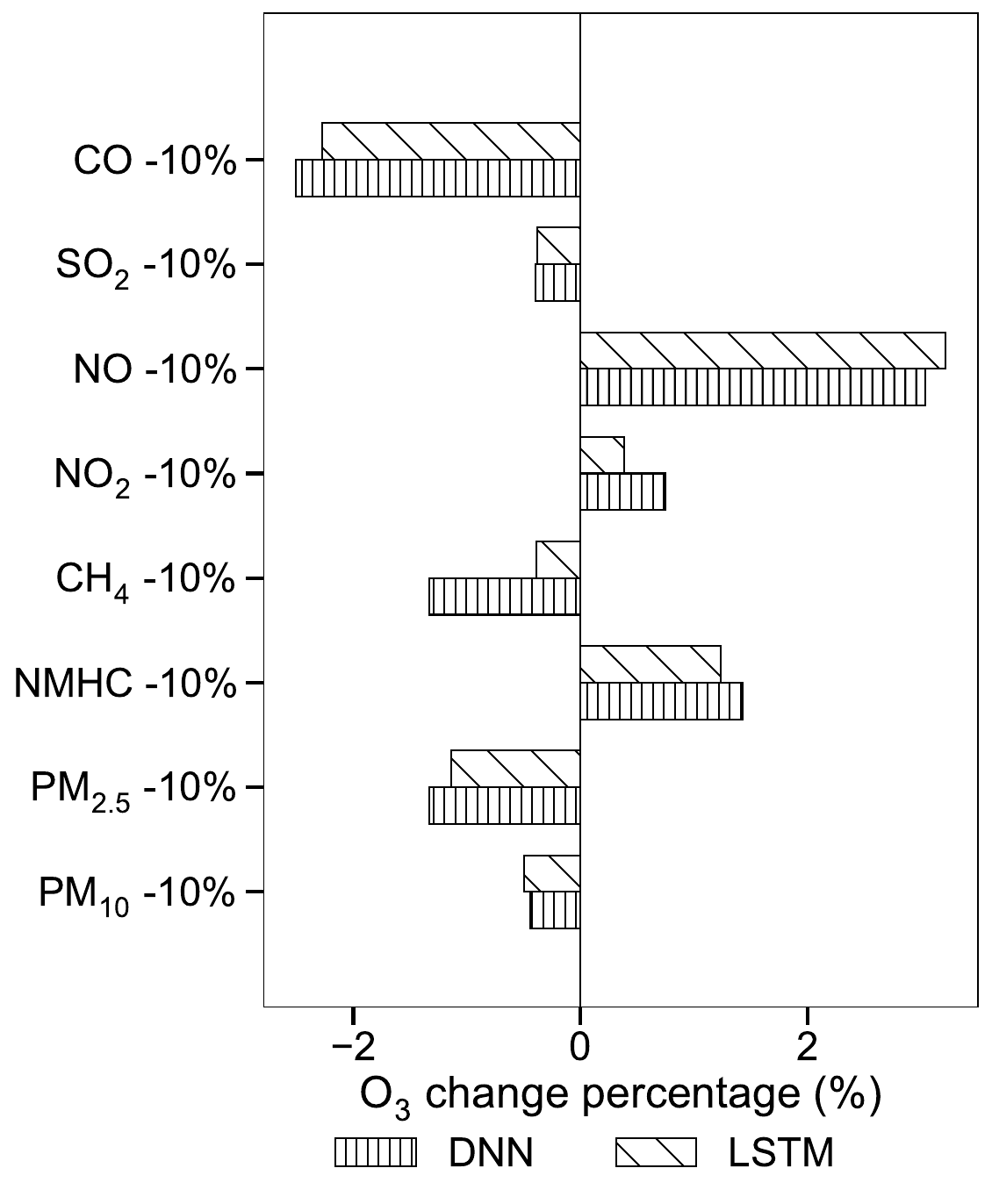} \label{fig:reducing_simulation}}
    
    \caption{The predicted \ce{O3} change under (a) climate change scenarios and (b) pollution controlling scenarios. (a) Significant increase in temperature and water vapor content in the different CESM2 simulation based on four SSP assumptions is considered and applied our test dataset in DNN and LSTM models. (b) Besides, we consider the reduction of 10\% of each anthropogenic variables in the test dataset in DNN and LSTM models.}
   \vspace{-2mm} 
\end{figure*}

\subsubsection{Simulation of climate change and pollution reduction}
\paragraph{Simulation of climate change} As advised by the model simulation results, we apply the mentioned temperature and water vapor content increase to our test data (the observation during 2018) separately and together with the remaining variables unchanged and predict their effect on \ce{O3} concentration. Figure \ref{fig:cc_simulation} presents the results. The DNN and LSTM models both indicate that increasing temperature could raise \ce{O3} concentration while increasing water vapor content could lower \ce{O3} concentration. While the positive contribution of temperature increase could significantly raise \ce{O3}, the negative contribution of water content vapor increase is able to offset the effect of temperature increase. Consequently, the change in \ce{O3} becomes negligible when considering the perturbation of these two variables together (as shown in Figure \ref{fig:cc_simulation}).

\paragraph{Simulation of pollution reduction} The above results indicate that the reduction of anthropogenic pollutants might be more crucial for controlling \ce{O3} in the future. Reducing pollution is always an important policy move in most countries for improving public health in general. However, controlling different pollutants can have a distinct impact on \ce{O3} concentration. Thus, we study the effect of reducing 10\% of each anthropogenic pollutant value on the predicted \ce{O3} in test dataset applying DNN and LSTM models. Figure \ref{fig:reducing_simulation} shows the results which demonstrate that reducing 10 percent CO would have the most apparent effect on decreasing ground-level \ce{O3} among all anthropogenic variables found by both models. Again, though controlling the emission of CO could contribute to lower \ce{O3}, reducing the amount of \ce{NO} and \ce{NO2} might lead to an increment of \ce{O3}. Therefore, simulations with different pollution control strategies by global climate models are still necessary to have a more comprehensive evaluation. 

\paragraph{Discussion} As presented in Fig. \ref{fig:reducing_simulation}, reducing CO, \ce{CH4}, and $\rm PM_{2.5}$ could be important for decreasing \ce{O3} concentration. Anthropogenic CO comes from the incomplete combustion of carbon-based fuel, and major CO sources include transportation and industrial activity. The generation of \ce{CH4}, another major greenhouse gas, is also highly co-related to human activity, such as agriculture, fossil fuel extraction, wildfire, and biomass burning. $\rm PM_{2.5}$ are aerosols with complicated composition and can be directly emitted or formed via sophisticated chemical reactions of gases including \ce{NOx}, \ce{SO2}, and VOCs. To reduce CO, \ce{CH4}, and $\rm PM_{2.5}$, it will be important to decrease the use of fuel vehicles and carbon-containing fuel and raise the percentage of renewable energy. However, reducing anthropogenic gases means that the concentrations of NO and \ce{NO2} would also decrease. The negative contribution of NO and \ce{NO2} must be carefully studied to evaluate the total effect of reducing anthropogenic gases to the future \ce{O3} concentration. These results clearly show the various kinds of actions where the government should have a stricter policy to make a better environment for the future.

%% file: 4_conclusion.tex
\section{Conclusions}
\label{sec:conclusion}
In this study, we have predicted tropospheric \ce{O3} which is one of the greenhouse gas and an influential ground-level air pollutant that can severely damage the environment. We have compared six methods to estimate the tropospheric \ce{O3} concentration and understand the importance of some meteorological variables, trace gases and pollutants in forming the ground-level \ce{O3}. The importance of solar radiation is emphasized in the best two models, DNN and LSTM models, which conform to the theoretical study. However, all \ce{NOx} and volatile organic compounds (VOCs) are presented to contribute negatively to \ce{O3} prediction, which contradicts the \ce{O3} and \ce{NOx}-VOCs relationship. This would promote a direction for future research about undiscovered \ce{O3} formation mechanisms. Moreover, the study regarding the importance of the variables or factors will lead to better policy makings to control the production of such materials or pollutants. We have further investigated the \ce{O3} concentration under different scenarios and shown that controlling anthropogenic gases, especially CO, could be critical for reducing \ce{O3} in the future considering the facts that the surface temperature and water vapor content may increase. Our findings clearly show the various kind of actions that the government should have stricter policies on, to make a better environment for the future. 